\documentstyle[11pt]{article}
\begin{document}
\pagenumbering{arabic}

\def\Bbb{\bf}

\begin{flushright}
\renewcommand{\textfraction}{0}
June 8th 1995\\
hep-th/9506171\\
PEG-06-95\\
\end{flushright}

\begin{center}
{\LARGE {\bf The Small Scale Structure of Space-Time:\\
                    A Bibliographical Review} }
\end{center}
\begin{center}
{\Large Phil Gibbs} \\
e-mail to philip.gibbs@eurocontrol.fr
\end{center}

\begin{abstract}
This essay is a tour around many of the lesser known pregeometric
models of physics, as well as the mainstream approaches to quantum
gravity, in search of common themes which may provide a glimpse of
the final theory which must lie behind them.
\end{abstract}

\pagebreak

\begin{center}
{\LARGE {\bf Introduction}}
\end{center}

In this review I will reflect on some of
the developments in quantum gravity which have emerged over the last
15 years. It is remarkable that these have advanced despite the
lack of any experimental input at anywhere near the relevant energies.
The theories are built on purely mathematical foundations, turning the
clock back past 2,500 years of scientific method to emulate the ancient
methodology of Plato. It might well be asked in view of this
if physicists are accomplishing anything more than constructing better
Platonic solids.

But if they were working on such shaky ground we might expect to see
a collection of mutually incompatible ideas emerging. What encourages us to
believe that something related to real physics is being studied is
that the various threads of development show signs of deep connections
suggesting that they could all turn out to be aspects of some
underlying unified theory.

Many theoretical physicists believe that to progress much further it
will be necessary to rethink our understanding of space-time. The 
4D manifold structure of general relativity does not seem adequate to
describe the kind of processes which are implicated in quantum gravity.
This poses a difficult puzzle. Can we be hopeful that the necessary
mathematics to describe physics beyond the Planck scale is within our
understanding?

We can at least be optimistic that experimental physics has more to
tell us. Future accelerators may find supersymmetry or something else
unexpected and lead us to
the correct unified theory at the GUT scale. It may then be possible
to demonstrate that this model is an unfalsifiable low energy limit
of some more fully unified model such as string theory and in the 
absence of other plausible models most people will accept its validity. 
But that is not the end of the story. String theory lacks predictive 
power at high energies where it becomes interesting. 
To go beyond this point it will be necessary to understand the
nature of space-time itself. Experiment is unlikely to provide
any direct help since the appropriate energies are way beyond reach.

The outstanding problem in theoretical physics today is to
uncover the mathematical origins of string theory which would explain
its enigmatic properties and allow us to solve it. I contend that the
key discovery is that the symmetry of space-time must be extended to
one which is event-symmetric. 

As well as giving potted reviews of the mainstream theories of Quantum 
Gravity,
I have collected together here a number of the better theoretical ideas 
which have been proposed as ways to go beyond our current understanding 
of the structure of space-time and replace it with some form of 
pregeometry. These ideas have been developed largely in isolation
and from a variety of motivations. At present they are at the fringe
of mainstream physics. I find however that many common ideas have
emerged and that considerable mathematical maturity has been introduced
into the field in recent years. 

20 years ago there were very few
physicists who studied the small scale structure of space-time.
Among those who did the names of Wheeler \cite{MiThWh73} and Finkelstein 
\cite{Fin94} stand out as two who independently conceived many of the 
foundation principles. They were then ahead of their time but now things have 
progressed. We are beginning to understand enough about quantum gravity
to gain the necessary physical insight and with new mathematical
tools such as quantum groups we may have what we need to tackle the
problem. Against the odds it now starts to look
as if we have a chance to reach some understanding of space-time
beyond the traditional continuum model.

An important part of this paper is the list of references which
includes many articles on discrete models of space-time as well as
many others which provide clues and motivation. With the aid of
the QSPIRES database in particular I have constructed a diverse
list although it is by no means complete. Sadly many of the more
speculative ideas seem only to have appeared as preprints which
are never published or have been presented at conferences
and it is difficult to trace the proceedings some years later on.
Hopefully the introduction of electronic pre-print archives will
enable such documents to remain accessible in future.
My treatment of the work in the references is necessarily shallow
since there is a larger quantity than might have been imagined.
The reader will have to consult the original articles in order
to properly appreciate their significance. 

A part of this essay is devoted to my own work on models
of event-symmetric space-time. These models demonstrate a way in which
string field theories might be formulated in a
geometric non-perturbative setting with all its symmetry manifest.
Topology change, mirror manifolds and duality in string theory all
suggest that diffeomorphism invariance is too limited a symmetry for
string theory. The event-symmetric theory may be part of the 
solution to this problem.

This paper grew out of my own notes in which I have tried to collect
the pieces of a global picture of the theory to which quantum gravity
research is leading. I am probably not the right person to provide
such a general review since my general understanding of these topics
remains fairly shallow. However, I have found a great deal of enjoyment in 
comparing my own ideas and motivations with those of others and hope 
that this document will help introduce some others
to research on the small-scale structure of space-time.


{\LARGE {\bf Acknowledgements}}

I would like to express my gratitude to those who have made the literature
accessible on the internet. Special thanks must go to
Paul Ginsparg for setting up the physics e-print archives at Los Alomos.
Without the success of such a service my research would not have been
possible. I am also deeply indebted to the librarians at
SLAC, DESY and CERN for making the databases such as SPIRES available
on the World Wide Web. My work has greatly 
benefited from the facilities of the references and postscript databases.

My thanks also to A. K. Trivedi and E. Prugovecki for providing references
to show that lattice theory and quantum space-time have earlier origins
than I thought.
 
A special thanks also to the Biblioteque Interuniversaire Physique Recherche 
at Jessieu for
permitting me free access to journals in more conventional form.

\pagebreak

\section*{Is Space-Time Discrete?}

At a seaport in the Aegean around the year 500BC two philosophers,
Leucippus and his student Democritus, pondered the idea that matter was made 
of indivisible units separated by void. 
Was it a remarkable piece of insight
or just a lucky guess? At the time there was certainly no 
compelling evidence for such a hypothesis.  Their belief in the atom was a
response to questions posed earlier by Parmenides and Zeno. 
Perhaps they were also 
inspired by the coarseness of natural materials like sand and stone.
Democritus extended the concept as far as it could go claiming that not just
matter, but everything else from colour to the human soul must also consist
of atoms \cite{Smi94}. 

The idea was subsequently surplanted by the very different philosophies of
Plato and Aristotle who believed that matter was infinitely divisible
and that nature was constructed from perfect symmetry and geometry.
It was not until the eighteenth and nineteenth centuries that the atomic
theory was resurrected by Dalton, Maxwell and Boltzman. This time they
had better scientific grounds to support their belief. They were able to 
explain pressure, heat and chemical reactions in terms of interactions 
between atoms and molecules. Despite this indirect evidence the majority of 
scientists disfavoured the theory until Einstein explained that Brownian
motion could be seen as direct experimental evidence of molecular motions.
But how far has modern physics gone towards the ideal of Democritus that
everything should be discrete?

The story of light parallels that of matter. It was Newton who first
championed the corpuscular theory of light but without good foundation.
Everything he had observed and much more was explained by Maxwell's
theory of Electromagnetism in terms of waves in continuous fields.
It was Planck's Law and the photoelectric effect which later upset the
continuous theory. These phenomena could best be explained in terms
of light quanta. Today we can detect the impact of individual
photons on CCD cameras even after they have travelled across most of
the observable universe from the earliest moments of galaxy formation.

Those who resisted the particle concepts had, nevertheless, some good sense. 
Light and matter, it turns out, are both particle and wave at the same time.
The paradox is resolved within the framework of Quantum Field Theories
where the duality arises from different choices of basis in the
Hilbert space of the wave function.

After matter and light, history is repeating itself for a third time
and now it is space-time which is threatened to be reduced to discrete
events. The idea that space-time could be discrete has been a recurring one in 
the scientific literature. A survey of just a few examples reveals that 
discrete space-time can actually mean many things and is motivated by a
variety of philosophical or theoretical influences.

It has been apparent since early times that there is something different
between
the mathematical properties of the real numbers and the quantities of
measurement in physics at small scales. Riemann himself remarked on this
disparity even as he constructed the formalism which would be used to
describe the space-time continuum for the next century of physics 
\cite{Rie76}. When you measure a distance or time interval you
can not declare the result to be rational or irrational no matter how
accurate you manage to be. Furthermore it appears that there is a limit
to the amount of detail contained in a volume of space. If we look under
a powerful microscope at a grain of dust we do not expect to see minuscule
universes supporting the complexity of life seen at larger scales. Structure
becomes simpler at smaller distances. Surely there must be some minimum
length at which the simplest elements of natural structure are found and
surely this must mean space-time is discrete.

This style of argument tends to
be convincing only to those who already believe the hypothesis. It will
not make many conversions. After all, the modern formalism of axiomatic
mathematics leaves no room for Zeno's paradox of Archiles and the tortoise.
However, such observations and the discovery 
of quantum theory with its discrete energy levels \cite{Boh28} and the
Heisenberg uncertainty principle \cite{Hei38} led physicists to speculate
that space-time itself may be discrete as early as the 1930's
\cite{AmIw30,Rua31}. In 1936
Einstein expressed the general feeling that {\it ``... perhaps the success
of the Heisenberg method points to a purely algebraic method of 
description of nature, that is, to the elimination of continuous functions
from physics. Then, however, we must give up, by principle, the space-time
continuum ...''} \cite{Ein36}. Heisenberg himself noted that physics must
have a fundamental length scale which together with Planck's constant and
the speed of light permit the derivation of particle masses \cite{Hei43,Hei57}.
Others also argued that it would represent a limit on the measurement of
space-time distance \cite{SaWi58}.
At the time it was thought that this length scale would be around $10^{-15}m$
corresponding to the masses of the heaviest particles known at the time
but searches for non-local effects in high energy particle collisions have
given negative results for scales down to about $10^{-19}m \cite{Kra95}$ 
and today the consensus is that it must correspond to the much smaller 
Planck length at $10^{-35}m$ \cite{Pla99}.

The belief in some new space-time structure at small length scales
was reinforced after the discovery of 
ultraviolet divergences in Quantum Field Theory. Even though
it was possible to perform accurate calculations by a process of
renormalisation \cite{Bet47,Schw48,Fey48,Dys52} 
many physicists felt that the method was incomplete
and would break down at smaller length scales unless a natural cutoff
was introduced.

A technique which introduces such a minimum length into physics by
quantising space-time was attempted by Snyder in 1947 \cite{Sny47a,Sny47b}. 
Snyder introduced non-commutative operators for space-time coordinates.
These operators have a discrete spectrum and so lead to a discrete
interpretation of space-time. The model was Lorentz invariant but
failed to preserve translation invariance.
Similar methods have been tried by others since 
\cite{Yan47,Fli48,PiUh50,HeTa54,Ich81,Ban85,Mad92,Gad94}. 
The quantisation results in differential operators being replaced by finite 
difference operators as if they were acting on a discrete space-time. 
The momentum space is therefore
compact. An alternative way to get a similar effect is to start from a
momentum
space which has a constant curvature \cite{Gol62,Cic66}. Although
no complete theory has come of these ideas there has been a recent upsurge of 
renewed interest in quantised space-time, now reexamined in the light of
quantum groups. We will return to this later. Another modern approach
to quantised space-time is provided by Prugovecki \cite{Pru94,Pru95}.

Quite a number of alternatives and variations on quantised space-time have
been 
tried over the years. Yukawa and Heisenberg and others considered non-local
models 
or field theory of particles which were not pointlike 
\cite{Hei43,Yuk50a,Yuk50b,Yuk53,Hei57,Ros73,Sog73}. 
Similarly again, the minimum
length can be introduced by stochastically averaging over a small
volume of space-time 
\cite{Zim62,Nam85,DiNa90,BrElSt95}. The bane of all these models is loss
of causality. We might regard superstring theory as the eventual successful 
culmination of this program \cite{Ich81} since it describes a field theory
of non-point like objects which respect causality.

Another way to provide a small distance cut-off in field theory is to
formulate it on a discrete lattice. This approach was also introduced early
by Wentzel \cite{Wen40} but only later studied in any depth
\cite{Das60,Bop67,Has72,Mee72,Wil74,Wel76}. Most recently lattice models
of space-time have been studied by Yamamoto et al 
\cite{Yam84,Yam85,YaHaHaHo93}
and Preparata et al \cite{PrXu91,PrXu94,Pre95}.
If the continuum limit is not to be restored 
by taking the limit where the lattice spacing goes to zero then the issue of 
the loss of Lorentz invariance must be addressed \cite{Lor74,Yam89}.
For some authors a space-time in which the coordinates take rational
values can be called discrete \cite{Hil55,HoKaKeUz92,NeSi93}. Lorentz 
invariance
is then possible but there is no minimum length scale.

None of these ideas were really very inventive in the way they saw
space-time. Only a rare few such as Finkelstein with 
his space-time code \cite{Fin69,Fin72a,Fin72b,Fin74,FiFrSu74} or Penrose with
twistor theory \cite{PeMa72} and spin networks \cite{Pen71}
could come up with any concrete suggestions for a more radical
pregeometry before the 1980's.

Another aspect of the quantum theory which caused disquiet was its
inherent indeterminacy and the essential role of the observer in 
measurements. The Copenhagen interpretation seemed inadequate and
alternative hidden variable theories were sought. It was felt that
quantum mechanics would be a statistical consequence of a more
profound discrete deterministic theory in the same sense that 
thermodynamics is a consequence of the kinetic gas theory.  

Over the years many of the problems which surrounded the development of
the quantum theory have diminished. Renormalisation itself has become
acceptable and is proven to be a consistent procedure in perturbation theory
of gauge field physics \cite{Hoo94a}. The perturbation series itself may not
be convergent but gauge theories can be regularised non-perturbatively on
a discrete lattice \cite{Wil74} and there is good reason to
believe that consistent Quantum Field Theory can be defined on continuous
space-time at least for non-abelian gauge theories which are asymptotically 
free \cite{Hoo82}. In Lattice QCD the lattice spacing can be taken to zero 
while the coupling constant is rescaled according to the renormalisation group 
\cite{Wil71}. In the continuum limit there are an infinite number of degrees 
of freedom in any volume no matter how small. The Nielsen-Ninomiya no-go 
theorem 
\cite{NiNi81a,NiNi81b} spells a problem for the inclusion of fermions but this 
too may be possible to resolve \cite{Tre93,Hoo94b}.

Quantum indeterminacy has also become an acceptable aspect of physics.
Everett's thesis which leads us to interpret quantum mechanics as a
realisation of many worlds \cite{Eve56} has been seen as a resolution of the 
measurement problem for much of the physics community.

Without the physical motivation discrete space-time is disfavoured by many.
Hawking says {\it``Although there have been suggestions that space-time may
have a discrete structure I see no reason to abandon the continuum theories
that have been so successful''} \cite{Haw94}. Hawking makes a valid point
but it may be possible to satisfy everyone by invoking
a discrete structure of space-time without abandoning the continuum theories
if the discrete-continuum duality can be resolved as it was for light and 
matter.


\section*{Discreteness in Quantum Gravity}

It is only when we try to include gravity in Quantum Field Theory that we
find solid reason to believe in discrete space-time. With quantisation of
gravity
all the old renormalisation issues return and many new problems arise
\cite{Ish93}.
Whichever approach to quantum gravity is taken the conclusion seems to be that
the Planck length is a minimum size beyond which the Heisenberg Uncertainty
Principle prevents measurement if applied to the metric
field of Einstein Gravity \cite{Gar94}. This can be expressed in a generalised
uncertainty principle \cite{KoPaPr90,Mag93}.

Does this mean that
space-time is discrete at such scales with only a finite number of degrees
of freedom per unit volume? Recent
theoretical results from String Theories and the Loop-representation of 
Gravity do suggest that space-time has some discrete aspects
at the Planck scale \cite{KlSu88,KiYa84,AsRoSm92,RoSm94}. 

The far reaching work of Bekenstein and Hawking on black hole thermodynamics
\cite{Bek72,Bek73,Haw75,HaHa76} has led to some of the most compelling evidence
for discreteness at the Planck scale. The {\it black hole information loss 
paradox} \cite{Haw76} which arises from semi-classical treatments of quantum 
gravity is the nearest thing physicists have to an experimental result in 
quantum gravity. Its resolution is likely to say something useful about a more 
complete quantum gravity theory. There are several proposed ways in which the 
paradox may be resolved most of which imply some problematical breakdown of 
quantum mechanics \cite{DaSc93} while others lead to seemly bizarre 
conclusions.

One approach is to suppose that no more information goes
in than can be displayed on the event horizon and that it comes back out
as the black hole evaporates by Hawking radiation. Bekenstein has shown that 
if this is the case then very strict and counter-intuitive limits
must be placed on the maximum amount of information held in a region of space
\cite{Bek93,Bek94}. It has been argued by 't Hooft that this finiteness of
entropy and information in a black-hole is also evidence for the discreteness
of space-time \cite{Hoo85}. In fact the number of degrees of freedom
must be given by the area in Planck units of a surface surrounding the region 
of
space. This has led to some speculative ideas about how quantum gravity 
theories might work through a hologramic mechanism \cite{Hoo93,Sus94}, i.e.
it is suggested that physics must be formulated with degrees of freedom
distributed on a two dimensional surface with the third spatial dimension
being dynamically generated.

At this point it may be appropriate to discuss the prospects for 
experimental results in quantum gravity and small scale space-time
structure. Over the past twenty years or more, experimental high energy
physics has mostly served to verify the correctness of the Standard Model as
proposed theoretically in 1967 \cite{Wei67}. We now have theories 
extending to energies way beyond current accelerator technology but
it should not be forgotten that limits set by experiment have helped
to narrow down the possibilities and will presumably continue to do so.

It may seem that there is very little hope of any experimental input
into quantum gravity research because the Planck energy is so far
beyond reach. However, a theory of quantum gravity would almost certainly
have low energy consequences which may be in reach of future experiments.
The discovery of supersymmetry, for example, would have significant
consequences for theoretical research on space-time structure.

It seems unlikely that any experiment short of studying the death
throws of a
small black hole in the laboratory can give direct support for quantum
gravity research or fine structure of space-time.  
A number of possible signatures of quantum gravity have been identified
\cite{Smo95} and there is, controversially, some hope that effects may be
observable in realistic experiments see e.g. 
\cite{Ste85,GoHuNi86,Ant90,DeKuCa93,Khu94,AnBeQu94}.


\section*{It from Bit and the Theory of Theories}

In the late 1970's the increasing power of computers seems to have been
the inspiration behind some new discrete thinking in physics. Monte Carlo
simulations of lattice field theories were found to give useful numerical
results with surprisingly few degrees of freedom where analytic methods
have made only limited progress.

Cellular automata (see \cite{Wol86}) became popular at the same time with
Conway's invention of the Game of Life. Despite its simple rules defined on a
discrete lattice of cells the game has some features in common with the laws 
of physics. There is a maximum speed for causal propagation which plays a 
role similar to the speed of light in special relativity. Even more 
intriguing is the accidental appearance of various species of ``glider'' 
which move through the lattice at fixed speeds. These
could be compared with elementary particles. 

For those seeking to reduce physics to simple deterministic laws this was an 
inspiration to look for cellular automata as toy models of particle physics
despite the obvious flaw that they broke space-time symmetries
\cite{Fey82,Min82,Zei82,Tof82,Svo86}. Nevertheless the quest is not 
completely hopeless. With some reflection it is realised that a simulation 
of an Ising model \cite{Len20,Isi25} with a metropolis algorithm is a cellular 
automaton if the
definition is relaxed to allow probabilistic transitions. The ising model has
a continuum limit in which rotational symmetry is restored. It is important
to our understanding of integrable quantum field theories in two dimensions.
Other extended cellular automata can be used to model fermions \cite{Bia93}.
t'Hooft has also looked to cellular automata as a model of discrete 
space-time physics \cite{Hoo88,Hoo90,HooIsKa92,Hoo93}. His motivation is
somewhat different since indeterminacy in quantum mechanics is, for him,
quite acceptable. He suggests that the states of a cellular automata could
be seen as the basis of a Hilbert space on which quantum mechanics is
formed.

The influence of computers in physics runs to deeper theories than 
cellular automata.
There is a school of thought which believes that the laws of physics
will ultimately be understood as being a result from information theory. 
The basic unit of information is the binary digit or bit and the
number of bits of information in a physical system is related to its
entropy.

J.A Wheeler \cite{Whe90} has sought to extend this idea, {\sl ``.. every
physical quantity, every it, derives its ultimate significance from bits, 
a conclusion which we epitomise in the phrase, It from Bit''}. For 
Wheeler and his followers the continuum is a myth, but he goes further
than just making space-time discrete. Space-time itself, he argues, must 
be understood in terms of a more fundamental {\it pregeometry}
\cite{Whe80,Whe84,Whe94}.
In the pregeometry there would be no direct concepts of dimension or 
causality. Such things would only appear as emergent properties in the
space-time idealisation. 

The pregeometry precept rings true to many
physicists and even underlies many attempts to understand the deeper
origins of string theory. As Green puts it {\it ``In string theory there 
aren't four or ten dimensions. That's only an approximation. In the deeper
formulation of the theory the whole notion of what we mean by a dimension of 
space-time will have to be altered.''} \cite{DaBr88}.

Wheeler gives just a few clues as to how a pregeometry might be formulated 
of which the most concrete is the principle that the boundary of a boundary
is zero \cite{KhWh86}. This is a central result from algebraic topology
which has become significant in non-commutative geometry. 

The history of theoretical Physics has been a succession of reductions to 
lower levels, more fundamental, more unified, more symmetrical and ideally
simpler. There is a strong belief that this process must eventually finish
\cite{Wei93}
but with what? According to {\it It from Bit} the process will bottom out in 
some principle of information theory. A skeptic would demand how a single
mathematical principle  could be so important as to spontaneously bring about
the existence of the universe while others fail to do so. And how are we to
explain the importance of symmetry and the unreasonable effectiveness of
mathematics in physics as demanded by Wigner \cite{Wig60}?

The answers may lie in an understanding of algorithmic complexity. For
centuries mathematicians have looked at specific structures with simple
definitions and interesting behaviour. With the advent of powerful
computers they are beginning to look at general behaviour of complex
systems. It was Feigenbaum who made the discovery that complex systems
of chaotic non-linear equations exhibit a universal renormalisation 
behaviour characterised by the Feigenbaum constants \cite{Fei78,Fei79}. 
This type of universality has an independent existence which transcends 
details of the specific equations which generate it.

If we wish to understand the origins and meaning of physical law we may 
need to recognise them as the universal behaviour of a very general
class of complex systems. Algorithmic information theory is perhaps a
good place to look because computability is a universal property 
independent of the programming language syntax used to define it. In
the statistical physics of systems with a large number of degrees of
freedom we also find that the laws of thermodynamics emerge as a 
universal behaviour independent of microscopic details. Could we
apply statistical methods to the general behaviour of algorithms?

We know from Feynman's Path Integral formulation of quantum mechanics
that the evolution of the universe can be understood as a supposition
of all possible histories that it can follow classically. The 
expectation values of observables are dominated by a small subset of
possibilities whose contributions are reinforced by constructive
interference. The same principle is at work in statistical physics
where a vast state space is dominated by contributions at maximum
entropy leading to thermodynamic behaviour. We might well ask if
the same can be applied to mathematical systems in
general to reveal the laws of physics as a universal behaviour which
dominates the space of all possible theories and which transcends
details of the construction of individual theories. If this was the
case then we would expect the most fundamental laws of physics to have
many independent
formulations with no one of them standing out as the simplest. This
might be able to explain why such a large subset of mathematics is
so important in physics.

This philosophy is known as the Theory of all Theories. In general it
is rather hard to make progress since it would be necessary to
define an appropriate topology and measure in the space of all theories. 
It should
be possible to get away with some reasonable subset of theories which
forms a dense covering of the topology so that it has, in some sense,
arbitrarily good approximations to any significant point. In a
restricted form where the subset comprising all 2 dimensional 
conformal field theories is taken, there has been some qualified
success in understanding the non-perturbative origins of string theories
\cite{Wit92b,Hat93,RaSo93,Zwi93}. It is found that the renormalisation group
flow in the space of theories converges at fixed points which may indicate
the true vacuum of string theory.

Can we use the Theory of all Theories to explain why symmetry is so
important in physics? There is a partial answer to this question which 
derives from an understanding of critical behaviour in statistical physics.
Consider a lattice approximation to a Yang-Mills quantum field theory
in the Euclidean sector. The Wilson discretisation preserves a discrete
form of the gauge symmetry but destroys the space-time rotational symmetry.
If we had more carelessly picked a discretisation scheme we would expect 
to break all the symmetry. We can imagine a space of discrete
theories around
the Yang-Mills theory for which symmetry is lost at almost all points.
The symmetric continuum theory exists at a critical point in this space.
As the critical point is approached correlation lengths grow and details 
of the discretisation are lost. Symmetries are perfectly restored in the
limit, and details of all the different discretisations are washed out.

If this is the case then it seems that the critical point is surrounded
by a very high density of points in the space of theories. This is exactly
what we would expect if universal behaviour dominating in theory space
was to exhibit high symmetry. It also suggests that a dominant theory
could be reformulated in many equivalent ways without any one particular
formulation being evidently more fundamentally correct than another.
Perhaps ultimately there is an explanation for the unreasonable
effectiveness of mathematics in physics contained in this 
philosophy. If physics springs in such a fashion from all of mathematics
then it seems likely that discovery of these laws will answer many old
mathematical puzzles. Other similar arguments about origins of symmetry
can be found in the work of Nielsen et al \cite{BeBrNi87,FrNi91}.

The reader should be alerted to the fact that these arguments are at
best incomplete. The aim is to present a philosophical viewpoint which 
enables us to see that there could be some fundamental principle from which 
the laws of physics are derivable. If we except this heuristic argument then
there are two ways we can proceed. We could start by analysing the theory
space of very general complex mathematical systems in an attempt to
find the universal behaviour which dominates, or alternatively we can
look for possible pregeometries of space-time with high symmetry. 
In any case the Theory of Theories philosophy helps to widen our
horizons. It seems appropriate to leave behind the continuum nature
of space-time in the search for something more fundamental.


\section*{A Taxonomy of Pregeometries}

Those who choose to pursue Wheeler's idea that the geometry of space-time must
be replaced with a more fundamental pregeometry are faced with a difficult
task. They can not bank on any direct guidance from experimental 
results if the pregeometry structure reveals itself only at Planck scales.
If discreteness is an aspect of quantum gravity then we should be looking
for models which allow curvature of space-time. The quantised space-time
and regular lattice models are simply too rigid for this purpose. What then
should we choose as our guiding star to lead us towards a good theory.

There are a number of ways the problem might be approached theoretically. We
could study the properties of string theory, canonical quantisation of gravity
and the quantum gravity aspects
of black holes in the hope that this would lead us to a mathematical discovery
which reveals the true nature of space-time to us. This is certainly being 
done by some of the best theoreticians in the business and has offered many
important clues. A second approach for the more philosophically inclined is to
try and deduce physics from basic principles such as a theory of theories
or algorithmic information theory. Perhaps that is too ambitious since it
has rarely proven to be the case in the past that physical law shares our 
philosophical preferences.

We might instead start to look at a variety of possible mathematical models
in the hope that some model will be found to have properties which fit in
with what we think we are looking for. Perhaps we can take the best ideas
which have been tried in the past and crossbreed them in the hope of forming
better models which combine the best qualities of their parents. There have 
indeed been many speculative models described in the literature and it is
one of the aims of this paper to provide a large bibliography of references
which the reader might use as a kind of gene pool for which to breed new
models. The fact that many models have common features some of which are
mathematically quite sophisticated suggests that it may really be possible
to put together a theory which works.

To be a little more systematic it might be useful to make a list of some
of the properties and features of space-time and dynamics in conventional
physics 
theories. Some of these will need to be discarded in the formulation of our
pregeometry models and we can then feel encouraged if they re-emerge as a
dynamically induced property of the theory either in an exact or approximate
form.

{\underline{ Continuity: } A feature of space-time in both general 
relativity
and particle physics is its continuity. Space-time is modelled as a manifold
with continuous coordinates. We have already argued that space-time may be
discrete in some sense. Should we then start from a model without continuous
coordinates and look for them to reappear in some form? Coordinates certainly
have an artificial flavour even in the mathematical description of a manifold.
Does this mean they have to be replaced with some discrete structure or should
the discreteness be derived from a continuous model in the same sense that an
atom has discrete energy levels which can be derived from the Shrodinger 
equation?}

{\underline{ Dimension: } Physicists have from time to time tried to
answer the question as to why space-time has 3+1 dimensions 
\cite{Ehr17,Whi55,Vol89,HoWh91,NiRu94}. It is well known that supergravity
and superstring theories are most consistent in 10 dimensions and that
the observed number of dimensions must come about through compactification 
of 6 spatial dimensions if such a theory is correct. Some authors consider 
the possibility that the number of dimensions could actually change in
a phase transition at high energies 
\cite{AhImMa76,LiZe88,BlGu89,Ant90,Pel91}. If space-time has a fractal
structure then the number of dimensions can be variable and non-integer
\cite{Svo87,Eyi89}.
Should we therefore abandon dimensionality as fundamental altogether and
start from a model which has no pre-set dimension? If space-time is discrete
then it is possible that the number of dimensions can be derived dynamically
and a number of pregeometry models of this type have
been attempted 
\cite{DaPi79,KaWe85,Jou85,Jou86,Sor91,BrGr91,AlCeVe92,Ant94a,Gib95a,Req95}.}

{\underline{ Metric: } The space-time metric is the fundamental dynamical
field in Einstein's formulation of general relativity and so one approach
to generalising space-time is to look at more general geometric spaces
which have a distance function defined on them \cite{Alv88,AlCeVe92,IsKuRe90}.
General Relativity can,
however, be reformulated in alternative ways making the metric less
fundamental and already in string theory the metric tensor is just an
aspect of a spin-two field which is a consequence of the dynamics. It may
therefore be quite natural to discard the metric as a fundamental feature.
Sometimes the term ``pregeometric'' is used to describe field theory defined
on continuous manifolds without a metric structure, including Topological
Field 
Theory \cite{Aka81,Aka84,FlPe90,Ter91,AkOd91,AkOd93}. Many of the more 
interesting
pregeometry models encountered in this review discard both the metric and 
continuity}

{\underline{ Causality: } Loss of fundamental causality in a pregeometric 
model is difficult for some physicists to come to terms with. One pregeometric
approach is to treat causality as the most important feature of the theory.
This is typified by models of physics on causal structures
\cite{Fin69,BoLeMeSo87,Fin88,Fin89,Sta91,BrGr91,Sza92}. On the other hand some 
physicists
do not see causality as a fundamental part of physics but rather as something
which is guaranteed only by dynamical effects as in Hawking's Chronological
Protection Conjecture \cite{Haw92}. Causality could be violated at microscopic
scales or even macroscopically and whether or not it is might perhaps be 
better 
left as a consequence of the theory rather than a fundamental principle. The
most dramatic example of causality violation may be the Big Bang if it came
about without any cause.}

{\underline{ Topology: } Wheeler emphasised the importance of topology
in the small scale structure of space-time \cite{Whe62}. 
There are close ties between causality and topology
on space-time since breakdown of causality may be linked to closed time-like
curves and wormholes. A number of important papers have studied how to
derive topology from discrete structures such as graphs or
partially ordered sets (posets) 
\cite{Ish89,IsKuRe90,Sor91,BaBiErTe93,Zap93,Ane94,Eva94,Zap95b}.
It is possible that a pregeometry may have no 
exact representation of topology and that it only arises dynamically or
perhaps in a way which is less direct. An example of topological theory
which may be connected with quantum gravity and space-time topology
is knot theory \cite{RoSm88,Wit89,Ati90,Kau91,Bae92}. Knot theory is
automatically brought into
physics whenever quantum groups are used and it is possibly through
higher-dimensional algebras that topology may be understood in 
pregeometry \cite{BaDo95}.}

{\underline{ Symmetry: } The importance of symmetry in physics is often
stressed. Much theoretical research has proceeded on the assumption that
the symmetries so far observed in nature are just part of a larger symmetry,
most of which is hidden by spontaneous symmetry 
breaking at low energy. Despite this, very few discrete pregeometries
have been able to reflect space-time symmetries in a discrete form and
some have suggested that such symmetry is not fundamental or even not
exact \cite{NiPi83,Yam89}. There are, however, a small number of
possible approaches which do consider space-time symmetries in a
discrete framework: Lattice Topological Field Theories
\cite{FuHoKa92,ChFuSh93}, Event-symmetric Space-time, \cite{Gib95a} 
and non-commutative geometries with quantum group symmetry
\cite{Con86,Con87}.}

{Internal gauge symmetry can be more easily represented in discrete 
form on a lattice as was demonstrated by Wilson \cite{Wil74}.
The key to finding a successful pregeometry theory may be to formulate
a unification of space-time and internal symmetries even if it is at
the cost of causality, continuity and topology. It should also be
born in mind that the concept of symmetry may have to be generalised
from group theory to include quantum groups and possibly other more
general algebras from category theory \cite{Cra93a,BaDo95}. Ultimately 
symmetry may be understood as a consequence of something more fundamental.
See for example \cite{FoNiNi80,FrNi91}.}

{\underline{ Quantum Mechanics: } There are some researchers who are 
strongly motivated by the desire to replace quantum mechanics with
something else. Usually they would prefer a deterministic theory.
Despite all the effort and many years of debate nobody has produced
an experiment which is not in agreement with the quantum theory
and most physicists regard its interpretation as being at 
least separated from physical dynamics.} 

{But what form should quantum mechanics take in pregeometry? The
answer depends on what kind of pregeometry you have and on what
formulation of quantum mechanics you prefer. The main choice 
which must be made is whether to start from the 
canonical (Hamiltonian) formulation of quantum mechanics or from the path
integral (Lagrangian) formulation \cite{Fey48}. Other possibilities
exist such as a purely algebraic construction of an S-matrix.} 

{The major difficulty with
the canonical formulation is that it normally assumes a continuous
time variable. If time is discrete or not defined at all then some
modification of the procedure is needed \cite{BaCh94}. However, with
the canonical approach we at least get a well defined notion of state
and can control unitarity.}

{If the canonical approach cannot be used or if a symmetrical 
treatment of time and space is desired then the path integral
approach is better suited. A sum over histories can be defined in
a very general context provided it is only desired to define some
generalisation of the partition function or Greens functions. If
we wish to understand the concept of state or calculate transition
amplitudes then we are again in trouble. With the path integral
approach we can postpone questions of causality and unitarity till 
later or they can be tackled with some generalised formalism such as
that of Hartle \cite{Har93}.}

For each of the above, a decision must be made as to whether it should be 
regarded as a fundamental feature of the pregeometry or a dynamically
derived property with possibly only approximate validity. It seems to be
mostly a matter of taste which governs the course taken by each researcher
but at least it should be a consciously made choice. Perhaps there is not
really a right or wrong approach. It is quite possible that different
models based on different principles can turn out to be equivalent.

Once it has been decided which properties of space-time are to be discarded
and which are to remain fundamental, the next step is to choose the
appropriate geometric structures from which the pregeometry is to be
built. An obvious candidate for a fundamental building block is the
space-time event. A continuous space-time manifold is itself a set
of events with certain other structures built on top. It would be
natural to retain the set of events and then replace the structures
with something else. Wheeler described this as a "bucket of dust"
\cite{Whe64,MiThWh73}.

The space-time event does indeed appear as fundamental in many
pregeometry models together with various other structures. E.g.
a random graph can be composed of events and links which randomly
connect them. The importance of events is also emphasised in some
attempts to resolve the quantum measurement problem 
\cite{Whi29,Sta79,Haa90,BlJa94} and this field sometimes overlaps with
pregeometry models.

The causal net and poset models also use events as basic structures.
Other models are built on cell complexes \cite{Jou85,Jou86,Jou87,Van93}
or simplicial complexes \cite{Reg61} in which the event is just
a special case of a simplex. An event might also be a special case
of a discrete string.

Sometimes the geometric approach is abandoned in favour of logic
\cite{Fin69,GoGrWe92,Zap93,Ant94b} or algebra
\cite{Ger72,BaGi94,Ban94,Zap95a}.
Given a linear algebra we may take a basis of the algebra and associate
each component with a physical object such as an event, simplex or string.
However, a choice of basis is not unique and is only secondary to the
algebraic structure, so the reality of those objects
is subjective. There may be a group of transformations which 
generate changes of basis under which the theory is invariant. Perhaps
physical objects become real only through spontaneous breaking of
such symmetries.

An algebraic approach which has received a great deal of attention is 
non-commutative geometry \cite{Con86,Con87,CoLo91}. The topological structure
of space-time can be understood in terms of the differential algebra of
functions on the manifold. According to Connes it is natural to 
generalise space-time by using other commutative algebras or 
non-commutative algebras and differential forms. The differential
operator $d$ has the property that $d^2 = 0$. This is an algebraic
embodiment of Wheelers boundary of a boundary principle \cite{KhWh86}.

If the differential calculus on the product of a space-time manifold
and a discrete space is constructed, it is possible to recover the
standard model of electro-weak interaction with the Higgs field
appearing as the connection on the discrete space 
\cite{BaGuWa91,MoOk94} and more general unified models can
be constructed from different spaces \cite{ScZy93}. There are by now
many references on this subject.

An obvious generalisation as a way to unify the particle forces 
with gravity is to look at noncommutative geometries or differential
calculus on discrete sets or groups which represent the events
and symmetries of space-time  \cite{DiMu94a,DiMu94c,DiMuVa94}.
There is an important correspondence between such geometries and 
topological pregeometries formulated on posets \cite{DiMu94b,BaBiErLa94}

Finally, the importance of spin-structure in pregeometry must be noted.
Wheeler points out that if the topology of space-time is non-trivial
then spin-1/2 fields must be accounted for \cite{Whe62,MiThWh73}. A number of
pregeometry models have used spin structure in inventive ways
\cite{Pen71,PoRe68,Har90,FiGi93}.

For the remainder of this review I will look at what some of the major
approaches to quantum gravity have to tell us about the nature of 
space-time at small distances.


\section*{Lattice Models of Gravity}

Following the success of perturbative quantum field theory when applied to
the electromagnetic and nuclear interactions it was natural for particle
physicists to try the same approach to quantising gravity. The result was
a disastrous theory in which covariance was lost and renormalisability
could not be achieved \cite{DeNiBo75,GoSa86}. Quantum Gravity research remained
at this dead end for some time until it was realised that gravity was
somewhat different from other forces. Weinberg has shown that quantum
gravity could be finite if it has a suitable ultraviolet fixed point
\cite{Wei80}.

QCD, the theory of the strong interaction can be 
formulated and analysed using Lattice Gauge Theories. The Wilson
plaquette loop action for QCD \cite{Wil74} on a lattice formulated 
in the Euclidean sector 
has a certain elegance since it preserves a discrete version of gauge
invariance and uses a group representation rather than a representation of a
Lie algebra. Translation invariance is also preserved in a discrete form
but the rotation group is only coarsely represented.

It is natural to ask whether a useful non-perturbative
formulation of quantum gravity can be found on a lattice as it can for QCD. 
The immediate objection to this is that a
lattice theory cannot reflect the important symmetries of space-time
in a discrete form. It seems that they can only be recovered in the
continuum limit.

It is common practice in Lattice Gauge Theories to convert the quantum theory
into a 4 dimensional Euclidean statistical physics by a Wick rotation from real
time to imaginary time. In the Euclidean sector numerical studies are feasible
in Monte Carlo simulations.

Lattice studies of gravity are likewise made typically in a Riemannian sector
where the metric has an all positive signature. 
A Riemannian theory of gravity would, however, present some
interpretation problems since it is not possible to simply apply a
Wick Rotation as can be done for the Euclidean sector of
lattice theories without gravity. Nevertheless, if it could be shown that there
was a continuum limit this would be strong evidence for the existence of a
quantum gravity theory without the need to extend the classical theory.

Lattice studies of pure gravity start from the Regge Calculus 
\cite{Reg61} in which space-time is ``triangulated'' into a simplicial complex.
The dynamical variables are the edge lengths of the simplices.
In 4 dimensions an action which reduces to the usual Einstein Hilbert action 
in the
continuum limit can be defined as a sum over hinges in terms of facet areas
$A_h$ and deficit angles $\delta_h$ which can be expressed in terms of the 
edge lengths.
\begin{equation}
                  S = \sum_h k A_h \delta_h
\end{equation}
The model can be studied as a quantised system and
this approach has had some limited success in numerical studies 
\cite{RoWi81,RoWi84,Ham85,Ham91}.

One way to retain a form of diffeomorphism invariance is to use a random
lattice on space-time instead of a regular lattice \cite{ChFrLe82,Lee83}.
A random lattice does not prefer any direction and Poincare invariance
can be exact.
An interesting aspect of the use of random lattices for quantum
gravity is that if the lattice is allowed to change dynamically in
some probabilistic fashion then its fluctuations merge with the
quantum mechanics in the Reimannian sector. An action with fixed edge
lengths but random triangulations \cite{BoKaKoMi86} is given by,
\begin{equation}
                 S = - \kappa_4 N_4 + \kappa_0 N_0 
\end{equation}
The partition function is formed from a sum over all possible triangulations
of the four-sphere. $N_4$ is the number of four simplices in the triangulation
and $N_0$ is the number of vertices. The constant $\kappa_4$ is essentially
the cosmological constant while $\kappa_0$ is the gravitational coupling 
constant. Random triangulations
of space-time appear to work somewhat better than the Regge Calculus with
a fixed triangulation.

4 dimensions is qualitatively different from 2 or 3 dimensions
where there can be no gravitational waves and hence no gravitons. 
In 4 dimensions there is evidence from numerical simulations that there 
is a second order phase transition under variation of the gravitational 
coupling constant and that a continuum limit with the correct Hausdorff
dimension can be found at the critical point \cite{AgMi92}. There is some
debate about whether the number of triangulations is exponentially bounded
\cite{CaKoRe94a,CaKoRe94b,AmJu94}.
This is an important condition for the model to be well defined in the
large volume limit. It will probably require numerical studies on quite large 
lattices to settle this issue.

The question of the exponential bound is extremely important in quantum 
gravity. If it is not there then that might be interpreted as evidence
that topology is important in the microscopic view of space-time. The
models of Numerical triangulations studied numerically are restricted to
a simple closed sphere topology. If this needs to be extended to a sum over
other topologies to include contributions of wormholes and the like then
it is important to know what weight should be given to each topology.

Even if dynamical triangulations provide a useful calculation method for
quantum gravity there are some unanswered questions concerning its
suitability as a fundamental formulation, among these is the
question of ergodicity or computability \cite{NaBe92}.


\section*{String Theories}

Despite the lack of experimental data above the Electro-Weak energy scale, the 
search for unified theories of particle physics beyond the standard model has 
yielded many mathematical results based purely on constraints of high symmetry,
renormalisability and cancellation of anomalies. In particular, space-time 
supersymmetry \cite{WeZu74} has been found to improve perturbative 
behaviour and to bring the gravitational force into particle physics. One 
ambitious but popular line of quantum gravity research is superstring theory 
\cite{GrSc81a,GrSc81b,GrSc82}. String models were originally constructed 
in perturbative form and were found to be finite at each order \cite{Man92}
but incomplete in the sense that the perturbative series were not Borel 
summable \cite{GrPe88}.

Despite this there has been a huge amount of interest in a number
of super-string theories and in the Heterotic String in particular 
\cite{GrHaMaRo85}. The fact that this theory has an almost unique
formulation with the interesting gauge group $E_8 \otimes E_8$ persaudes
many that it is the sought after unified field theory despite the fact
that it is only finite in ten dimensional space-time. In 1926 Klein 
\cite{Kle26} proposed that a 5 dimensional theory due to Kaluza \cite{Kal21}
could make physical sense if one of the dimensions was compactified.
Kaluza-Klein theory has been applied to the heterotic string theory for which 
it has been shown \cite{CaHoStWi85} that six of the ten dimensions could be 
spontaneously compactified on a Calabi-Yau manifold or orbifold. This leaves 
an $E_6$ gauge group with suitable chirality modes just big enough to
accommodate low energy particle physics. The difficulty which remains
is that there are many topologically different ways the compactification 
could happen and there is no known way of picking the right one. To solve 
this problem it is thought necessary to find some non-perturbative analysis 
of the string theories. As a first step it might be necessary to construct 
a second quantised covariant String Field Theory \cite{KaKi74}.

There has been some preliminary success in formulating both open \cite{Wit86} 
and closed \cite{Zwi92} bosonic String Field Theories.
There have also been some important steps taken towards background independent 
formulations of these theories \cite{Wit92b,SeZw93}.
However, they still fail to provide an explicit unification of
space time diffeomorphism symmetry with the internal gauge symmetry. This is a
significant failure because string theories are supposed to unify gravity
with the other gauge forces and there is evidence that string theory does 
include such a unification \cite{HoHoSt91}. Furthermore these formulations 
have not yet been extended to superstring theories and this appears to be a 
very difficult problem \cite{Ced94}.

Conventional wisdom among the pioneers of string theories was that there 
is a unique string theory which is self consistent and which explains all
physics. This view was gradually tempered by the discovery of a variety of
different string theories but recently the belief in uniqueness has been
reaffirmed with the discovery that there are hierarchies in which some
string theories can be seen as contained within others 
\cite{BeVa93,BeFrWe93,KuSaTo94b}. This inspires a search for a universal
string theory \cite{Fig93,BaNoPe94,KuSaTo94a}.

A successful theory of Quantum Gravity should describe physics at the
Planck scale \cite{Pla99}. It is likely that there is a phase 
transition in string theories at their Hagedorn temperature near $kT =$ Planck
Energy \cite{Hag68}. It has been speculated that above 
this temperature there are fewer degrees of freedom and a restoration of a much
larger symmetry \cite{AtWi88,GrMe87,GrMe88,Gro88}. This phase is sometimes
known as the topological phase because it is believed that a Topological 
Quantum 
Field Theory may describe it. A fundamental formulation of
string theory would be a model in which the large symmetry is explicit.
It would reduce to the known formulations after spontaneous symmetry
breaking below the Hagedorn Temperature.

One interpretation of the present state of string theories is that it
lacks a geometric foundation and that this is an obstacle to finding
its most natural formulation. It is possible that our concept of 
space-time will have to be generalised to some form of ``stringy space''
in which its full symmetry is manifest. Such space-time must be 
dynamical and capable of undergoing topological or even dimensional changes 
\cite{AsGrMo93,AnFeKo94,GrMoSt95}. To understand stringy space it is almost 
certainly necessary to identify the symmetry which is restored at high 
temperature. 

The notion that string theory has a minimum length is well established as a
result of target space duality which provides a transformation from 
distances $R$ to distances $\alpha/R$ where $\sqrt \alpha$ is the size of 
compactified dimensions at the Planck scale. A minimum length does not
necessarily imply discrete space-time but it is suggestive.

Thorn has argued that large $N$ matrix models lead to an interpretation of 
string theories as composed of pointlike partons \cite{Tho94a,Tho94b,BeTh95}.
A similar view has been pursued by Susskind as a resolution of paradoxes
concerning Lorentz Contraction at high boosts and the black hole information 
loss puzzle \cite{Sus93,Sus94}. It is possible to calculate exact string 
amplitudes from a lattice theory with a non-zero spacing \cite{KlSu88}. 

There are some remarkable features about these discrete string models. Firstly 
it is found that when the spacing between discrete partons is reduced below
a certain limit there is a phase transition beyond which results coincide
exactly with continuous models \cite{KlSu88,Dal94,Kos94}. Even
more odd is the apparent generation of an extra dimension of space so
that models in 2+1 dimensions could become theories of 3+1 dimensions 
\cite{Rus93,Tho94a,SuGr94}. This has also been seen as an aspect of of the
continuum theory \cite{Wit95} and it seems that these concepts are not 
inconsistent with the algebraic construction of Topological Quantum Field
Theory \cite{Cra95}.

Perhaps the fact that string theory is finite to each order in perturbation 
theory is itself an indication that string theory is discrete. In lattice 
theories the renormalisation group is used to send the lattice spacing
to zero but in string theory the coupling is not renormalised.

If a string is to be regarded as made up of discrete partons then it
might make sense to consider the statistics of each parton. In the two
dimensional worldsheet of the string a parton could have fractional
statistics. If string partons are such that an increasing number of
them are seen in a string at higher energies it may be necessary for
the statistics to be divided up into fractions of ordinary fermionic
or bosonic statistics. In the higher dimensional target space only 
half integer multiple statistics are permitted to be observed.

Heuristically we might picture the string as an object consisting of
$n$ partons each with an interchange phase factor $q$ such that
$q^n$ is real, i.e. $1$ or $-1$. This suggests that a continuum limit
might exist where $n \rightarrow \infty$ on the worldsheet while the
string has discrete aspects in target space. Such a model might  
be based on quantum group symmetries. There are already some encouraging
results which suggest that it might be possible to formulate fractional
superstring models \cite{ArTy91}.

This section would not be complete without referring to a number of other
attempts to understand discrete string theory
\cite{GePa87,FiIs89,FiIs90,AlBa91,FiGaIs92}.


\section*{Canonical Quantisation of Gravity}

There has also been some progress in attempts to quantise Einstein
Gravity \cite{Ein15a,Ein15b} by canonical methods. A reformulation of 
the classical theory in which the connection takes the primary dynamic 
role instead of the metric \cite{Ash86,Ash87} has led
to the Loop Representation of Quantum Gravity \cite{RoSm88,RoSm90} in
which knot theory plays a central role.
The fact that Einstein Gravity is non-renormalisable is considered to be not 
necessarily disastrous since gravity theories in 1+1 and 2+1 dimensions have
been successfully quantised by various means \cite{Car93}. In the latest
versions of the theory the quantum gravity states are based on Penrose
spin networks \cite{RoSm95}.

String theory originated as a proposed theory of the strong interactions
before being replaced by QCD. It is possible that QCD may still be possible
to reformulate as a string theory at least approximately if not exactly.
The loop representation also first appeared in connection with Yang Mills
theories like QCD. \cite{GaTr86}

Another likeness between the loop representation and string theories is that
there are attempts to understand them in terms of field variables and groups
defined on loop objects \cite{GaTr81}. This and other similarities may be more 
than superficial \cite{Bae93}. Superstring theory and the Loop Representation 
of quantum gravity in the forms we know them can not be equivalent since the 
former only works in ten or eleven dimensions while the latter only works in 
three or four. It is possible that they could be different phases of the same 
pre-theory provided that pre-theory allows changes of dimension. Alternatively
it could just be that we just have not learnt how to do string theory in 4 
dimensions and the loop representation in 10 dimensional superspace yet,
or more likely that the role of dimension has been so far misunderstood.

A major difference between the canonical quantisation and string theory is
that it is an attempt to quantise gravity in the absence of matter fields.
The view from Canonical Quantisation
supports that of the lattice quantum gravity schemes that a theory of quantum
gravity without matter exists but
string theorists often express the belief that gravity can only be quantised
when other fields are included.

If quantum gravity was an easy business then somebody would have found the
formalism which allows you to form the perturbative expansion of the loop
representation about a fixed background and demonstrate its equivalence to
string theory. Differences between the two suggest that their relationship
may not be so simple and in fact the loop representation has at least as many 
consistency problems to resolve as string theory \cite{Ash94} before such 
a program might be realised.

Nevertheless the program is developing rapidly and the one of its great
strengths is that the form of the theory is derived directly from canonical
quantisation methods of gravity rather than being constructed ad hoc. The
fact that this leads to a discrete spectrum for volume and area operators
\cite{RoSm94} is a powerful argument in favour of discrete aspects of 
space-time in quantum gravity. In one version the theory is formulated on
a discrete lattice without losing diffeomorphism invariance enabling 
a convenient calculation scheme \cite{Lol95}.

Canonical quantisation of gravity may allow many other useful observations 
on the nature of space-time to be made.


\section*{Quantised Space-Time}

One of the most important principles in modern theoretical physics is
that of symmetry. It is quite likely that the observed symmetries in
nature are the remnants of a much larger symmetry which existed at
the beginning of the universe but which were successively broken down
to smaller and smaller symmetries as the universe expanded cooled. 
Knowing the full original symmetry is the key to knowing the laws of 
physics.

Traditionally the classification of symmetry was regarded as being
equivalent to the classification of abstract groups. This view changed
with the discovery of supersymmetry which allows us to define 
symmetries between fermionic and bosonic particles but which is not
related to a classical group. In the last decade a new type of symmetry
has been discovered and widely researched. This symmetry is related to an
algebraic object which is a deformation of the classical notion of group.
It is known as a quantum group. 

Quantum Groups first arose in the context of exactly solvable 1+1 dimensional
lattice systems \cite{Bax82} and the quantum inverse scattering method 
\cite{SkTaFa80}.
Drinfeld and Jimbo identified the relevant algebraic structure as a
quasi-triangular Hopf algebra \cite{Jim85,Dri86,Jim86}. The same structure
was discovered independently by Woronowicz in the context of 
non-commutative geometries \cite{Wor87,Wor89}. Since then quantum groups
have been recognised as important symmetries in many different types of
physical system and a large number of technical papers have been written.
In particular it is expected that quantum groups will have a major role
to play in theories of quantum gravity.

Quantum groups are an example of what mathematicians would call a 
deformation. A structure is defined which has a dependency on a complex
parameter $q$. In the special case $q=1$ the structure corresponds to
a group. In general it is not a group but still has many of the 
properties that make groups useful in physics.

Other types of deformation have already proved useful to physicists.
Quantum mechanics is a particular example which has a deformation
parameter given by Planck's constant $\hbar$. In the $\hbar \rightarrow 0$
limit we reach classical physics. Quantum groups may allow a new
deformation of physics which brings in a minimum length scale. It is
possible that this could lead to a deformed version of general relativity
which is finite when quantised. Another reason to suppose that quantum
groups are important in quantum gravity is the relationship between
quantum groups and knot invariants \cite{Kau91}. Knot theory is known to
be relevant to quantum gravity as topological field theory \cite{RoSm88}. 

For a detailed description of how a group can be deformed as a Hopf algebra 
there are many references which can be consulted 
\cite{Tji91,Fuc92,Rui93,Sch93,Koo94,Wat94}.

The concept of space-time quantisation goes back to the 1940's when
Snyder proposed that space-time coordinates should be replaced by
non-commutative operators \cite{Sny47a,Sny47b}. The aim was to introduce 
a fundamental length into physics in order to avoid divergences which 
plagued quantum field theory. (added note: The reader should consult
Sec. 1.3 of {\sl Principles of Quantum General Relativity} by 
E. Prugovecki \cite{Pru95} for even earlier references to papers on
quantised  space-time by Ruark, Flint, Richardson, Firth, Landau, 
Peierls, Glaser, and Sitte.)
Snyders model failed because although it was Lorentz invariant it 
destroyed translational invariance. Yang proposed a model on anti-de Sitter
space-time which also had translation invariance \cite{Yan47}. It was
unphysical because it implied a curvature of space-time on the
scale of the Planck length rather than the scale of the universe.
Townsend proposed a theory in which this could be realised by 
gaugeing the de Sitter group in place of the Poincare group \cite{Tow77}.

The problem of describing a discrete space-time
which has an adequate discretisation of the full Poincare group has
always been an obstacle to constructing field theory with
a fundamental minimum length scale. Schild deduced that an integer lattice
could preserve a large subgroup of the Lorentz group but one which was
far too coarse \cite{Schi48}. Hill argued that a space-time with rational
coordinates solved the problem \cite{Hil55} but it is debatable whether
such a dense covering can be described as discrete.

There are indeed some models of space-time which have some form of
space-time symmetry and a minimum length scale. One possibility is
to use a random lattice or dynamical triangulations. In this case 
symmetry is regained after quantisation which includes a sum over all
triangulations while keeping a minimum length scale. Again an event 
symmetric space-time also has a discrete version of diffeomorphism
invariance. We can also mention Lattice Topological Field Theory as
another example. 
These possibilities are described in other sections.

We may learn something useful if we can formulate a theory of
discrete space-time on flat space. The discovery of quantum groups has
brought about a revival of quantised space-time.
If our understanding of symmetry is broadened to include
quantum group symmetries then we can use q-deformed Lorentz \cite{PoWo90} or
Poincare groups \cite{LuRuNoTo91,PoWo94}. 
By factoring out the Lorentz algebra it is possible to define a deformation
of Minkowski space. Deformations of Euclidean spaces are equally possible
and worthy of study \cite{Fio94}. Many of these models have a discrete
differential calculus.

These spaces and groups are not defined directly. They are defined in terms
of an algebra of functions on the spaces. It is possible to construct
differential algebras which act on these functions and it is found that
these derivatives are finite difference operators. In this sense we have
succeeded in constructing discrete spaces without abandoning symmetry
principles.

The next step would be to construct field theories on these spaces. It
has been found that there are important constraints on which gauge
theories can be constructed 
\cite{ArVo91,BrMa92a,BrMa92b,IsPo93,ArAr93,Fin95}. 
It might also be interesting to try to gauge the symmetry in the same sense
as in gravity and supergravity theories \cite{Cas94}. If the flat model
is discrete then the gauged model which would describe curved space-time
could be especially interesting.

A limitation of quantum groups as a generalisation of symmetry is that
they are not general enough to include supersymmetry. The universal
enveloping algebra of a Lie superalgebra is not a Hopf algebra. It is
a super-Hopf algebra. An alternative structure developed by Majid is
the braided group which does include supersymmetry. Braided groups 
are related to quantum groups but perhaps the most general algebraic
structure to describe symmetry has not yet been defined. Majid also
uses his techniques to quantise space-time \cite{Maj94}.

There are many papers written on these and related subjects and it 
should also be noted that quantum groups and other related algebraic
structures are also of principle importance in string theory and
canonical quantisation of gravity.


\section*{Topological Quantum Field Theories}

A criticism often made against the way superstring theories have been
developed 
is that they are not explicitly covariant or background free. In contrast
Witten introduced the concept of Topological Field Theories and
represented gravity in three dimensions with the Chern-Simons-Witten (CSW)
model \cite{Wit88}. 
This turned out to have as many useful results in topology as it has in 
physics. Atiya considered how this would extend to quantum theories and
produced a set of axioms describing properties a quantum gravity should have
as a Topological Quantum Field Theory \cite{Ati90}.

A surprising series of discoveries which led to an alternative understanding
of Topological Field Theories had begun years before when it was found that
the Regge Calculus \cite{Reg61} in three dimensions could be approximated by 
a formula involving 6j-symbols from the quantum theory of angular momentum 
described by representations of the group $SU(2)$ \cite{PoRe68}. The 
Ponzano-Regge model is constructed from tetrahedral 
simplices having edge lengths which are quantised to half integer values.
These $j$-numbers have a dual interpretation as either spins or lengths. 
The relationship
with Penrose spin-networks \cite{Pen71} was also studied \cite{HaPe81}.

The Ponzano-Regge model
has now been reformulated as the Turaev-Viro model with SU(2) replaced by 
the quantum group $U_q(su(2))$ at q an r-th root of unity \cite{TuVi92}. This 
provides a natural regularisation of the model with the lengths limited to less
than $r/2$. The original formulation is recovered in the $q \rightarrow 1$ 
limit.
They showed that the partition function was independent of the triangulation
and could therefore only depend on the topology of the triangulated space. The
model is therefore a topological quantum field theory and can be regarded as
a successful quantisation of 3 dimensional gravity. 

It is now known that the Ponzano-Regge model is equivalent to a 
Chern-Simons-Witten model with gauge group $ISO(3)$ \cite{OoSa91,Oog92a} and
can also be transformed to a loop representation \cite{Rov93}.
Cherns-Simons models can be related to string theory \cite{Wit92a} and the
Turaev-Viro model can be reformulated in terms of surfaces (string worldsheets)
\cite{Iwa94} so within this 3-dimensional model we already see a unification
of many of the main-stream ideas in quantum gravity.

The equivalence between simplicial gravity and topological quantum field
theories could be interpreted as a partial resolution of the
discrete-continuous dual nature of space-time but some care is needed. 
In this duality the discreteness does not appear at just the Planck scale
unless space-time is assumed to be topologically complex at that scale in 
the sense of the space-time foam of wormhole geometrodynamics 
\cite{Whe57,Haw78}. In a topologically simple space-time the 
triangulation can be made so coarse that there would be only a few degrees 
of freedom left so there is a discreteness even at large scales. 
Furthermore it must be appreciated that three dimensional
gravity is very simple compared to four dimensional gravity, there are no local
excitations or gravity waves because there are no Weyl tensor components in 
less than four dimensions.

One further formulation of simplicial 3D gravity due to Boulatov 
is of particular interest. 
This starts from a perturbation theory
defined on a field of triangular objects moving on a quantum group. The 
triangular objects are made to interact through a tetrahedron vertex. The
perturbation expansion in the coupling constant of this pre-theory is then 
equivalent to the Turaev-Viro model \cite{Bou92a}.
This is of special interest because it is a pre-theory in which a 3 dimensional
space-time is dynamically generated and also because the interactions of the
triangular objects are reminiscent of string vertices in string field theory.

Given the success of this approach in 3 dimensions it is natural to try and
generalise to 4 dimensions and indeed several people have pursued this line
\cite{Oog92b,CrYe93,Bro93}. Sadly these models have turned out to be too 
simple to
be theories of quantum gravity so far \cite{CrKaYe93a,CrKaYe93b}. 
A more speculative proposal is that
4 dimensional physics can be found in the 3D simplicial models where 
tetrahedral inequalities are violated \cite{Laf93}. Such an approach would
be consistent with the hologram ideas of Thorn, t'Hooft and Susskind 
\cite{Tho94a,Hoo93,Sus94}. Another similar possibility is that the 3D TQFT
could be considered as a state of 4D gravity from which time evolution could
be inferred \cite{Cra93b}. Crane has been developing these ideas further in
order to try to understand the physical aspects to topological quantum field 
theories \cite{Cra95}.


\section*{Event-Symmetric Space-Time}

Finally I turn to my own research on space-time structure.

My belief is that the symmetry so far discovered in nature is just the
tiny tip of a very large iceberg most of which is hidden beneath a sea of
symmetry breaking. With the pregeometric theory of 
{\sl event-symmetric physics} I hope 
to unify the symmetry of space-time and internal gauge symmetry into one
huge symmetry. I hope that it may be possible to go even further than this.
Through dualities of the type being studied in string theory it may be
possible to include the permutation symmetry under exchange of identical 
particles into the same unified structure.

The theory of Event-Symmetric space-time is a discrete approach to quantum
gravity \cite{Gib95a}. The exact nature of space-time 
in this scheme will only become apparent in the solution. Even the number of 
space-time dimensions is not set by the formulation and must by a dynamic 
result. It is possible that space-time will preserve a discrete
nature at very small length scales. Quantum mechanics must be reduced to a
minimal form. The objective is to find a statistical or quantum definition of 
a partition function which reproduces a unified formulation of known and 
hypothesised symmetries in physics and then worry about states, observables
and causality later.

Suppose we seek to formulate a lattice theory of gravity
in which diffeomorphism invariance
takes a simple and explicit discrete form. At first glance it would seem that 
only translational invariance can be adequately represented in a discrete form 
on a regular lattice. Dynamical triangulation is much better but still the
symmetry is not explicit and only appears after quantisation.
This overlooks the most natural generalisation of 
diffeomorphism invariance in a discrete system. 

Diffeomorphism invariance 
requires that the action should be symmetric under any differentiable 
1-1 mapping on a $D$ dimensional manifold $M_D$. This is 
represented by the diffeomorphism group $diff(M_D)$. On a discrete 
space we could demand that the action is symmetric under any permutation 
of the discrete space-time events ignoring continuity altogether. Generally I 
use the term {\it Event-Symmetric} whenever an action has an invariance
under the Symmetric Group $S({\cal U})$ over a large or infinite 
set of ``events'' ${\cal U}$. The symmetric group is the group of all 
possible 1-1 mappings on the set of events with function composition as the
group multiplication. The cardinality of events on a manifold of any number 
of dimensions is $\aleph_1$. The number of dimensions and the topology of the
manifold is lost in an event-symmetric model since the symmetric groups for 
two sets of equal cardinality are isomorphic.

Event-symmetry is larger than the diffeomorphism invariance of continuum 
space-time. 
\begin{equation}
                 diff(M_D) \subset S(M_D) \simeq S(\aleph_1)
\end{equation}
If a continuum is to be restored then it seems that there must be a mechanism 
of spontaneous 
symmetry breaking in which event-symmetry is replaced by a residual 
diffeomorphism invariance. The mechanism will determine the 
number of dimensions of space. It is possible that a model could have several 
phases with different numbers of dimensions and may also have an unbroken  
event-symmetric phase. Strictly speaking we need to define what is meant by 
this type of symmetry breaking. This is difficult since there is no
order parameter which can make a qualitative distinction between a broken
and unbroken phase.

The symmetry breaking picture is not completely satisfactory because it
suggests that one topology is singled out and all others discarded by
the symmetry breaking mechanism but it would be preferred to retain the 
possibility of topology change in quantum gravity. It might be more accurate
to say that the event-symmetry is not broken. This may not seem to 
correspond to observation but notice that diffeomorphism invariance of
space-time is equally inevident at laboratory scales. Only the Poincare
invariance of space-time is easily seen. This is because transformations of
the metric must be included to make physics symmetric under general coordinate
changes. It is possible that some similar mechanism hides the event-symmetry.

It is possible to make an argument based on topology change that space-time
{\it must} be taken as event-symmetric in Quantum Gravity. Wheeler was the 
first to suggest that topology changes might be a feature of quantum 
geometrodynamics \cite{Whe57}. Over the past few years the arguments in 
favour of topology change in quantum gravity have strengthened see e.g.
\cite{BaBiMaSi95}. If we then ask what is the correct symmetry group in
a theory of quantum gravity under which the action is invariant, we
must answer that it contains the diffeomorphism group $diff(M)$ for any
manifold $M$ which has a permitted topology. Diffeomorphism groups are
very different for different topologies and the only reasonable way to
include $diff(M)$ for all $M$ within one group is to extend the group to
include the symmetric group $S(\aleph_1)$. There appears to be little
other option unless the role of space-time symmetry is to be abandoned
altogether.

It is unlikely that there would be any way to distinguish a space-time with an 
uncountable number of events from space-time with a dense covering of a 
countable number of events so it is acceptable to consider models in which the 
events can be labelled with positive integers. The symmetry group $S(\aleph_1)$
is replaced with $S(\aleph_0)$. In practice it may be necessary to regularise 
to a finite number of events $N$ with an $S(N)$ symmetry and take the large $N$
limit while scaling parameters of the model as functions of $N$.

Having abandoned diffeomorphisms we should ask whether there can remain
any useful meaning of topology on a manifold. A positive answer is
provided by considering discrete differential calculus on sets and finite
groups \cite{DiMu94a}.

In some of the more physically interesting
models the symmetry appears as a sub-group of a larger symmetry such as
the orthogonal group $O(N)$. It is also sufficient that the Alternating
group $A(N)$ be a symmetry of the system since it contains a smaller 
symmetric group.
\begin{equation}
                 S(N) \subset A(2N)
\end{equation}

The definition of the term event-symmetric could be relaxed 
to include systems with invariance under the action of a group which has
a homomorphism onto $S(N)$. This would include, for example, the braid 
group $B(N)$ and, of course, quantum groups such as $SL_q(N)$.

Renormalisation and the continuum limit must also be considered but it is not
clear what is necessary or desired as renormalisation behaviour. In 
asymptotically free quantum field theories with a
lattice formulation such as QCD it is normally assumed that a continuum limit 
exists where the lattice spacing tends to zero as the renormalisation group
is applied. In string theories, however, the theory is perturbatively finite
and the continuum limit of a discrete model cannot be reached with the
aid of renormalisation. It is possible that it is not necessary to have an
infinite density of events in space-time to have a continuum or there may
be some alternative way to reach it, via a q-deformed non-commutative geometry
for example.

It stretches the imagination to believe that a simple event-symmetric model 
could be responsible for the creation of continuum space-time and the 
complexity of quantum gravity through symmetry breaking, however, nature has 
provided some examples of similar mechanisms which may help us accept the 
plausibility of such a claim and provide a physical picture of what is
going on. 

Consider the way in which soap bubbles arise from a statistical physics 
model of molecular forces. The forces are functions of the relative 
positions and orientations of the soap and water molecules. The energy is
a function symmetric in the exchange of any two molecules of the same 
kind. The system is consistent with the definition of event-symmetry since it
is invariant under exchange of any two water or soap molecules and therefore 
has an $S(N) \otimes S(M)$ symmetry where $N$ and $M$ are the number of water 
and soap molecules. Under the right conditions the symmetry breaks 
spontaneously to leave a diffeomorphism invariance on a two dimensional 
manifold in which area of the bubble surface is minimised. 

Events in the soap bubble 
model correspond to molecules rather than space-time points. Nevertheless, it
is a perfect mathematical analogy of event-symmetric systems where the 
symmetry breaks in the Riemannian sector to leave diffeomorphism invariance in 
two dimensions as a residual symmetry. Indeed the model illustrates an
analogy between events in event-symmetric space-time and identical particles
in many-particle systems.
The models considered further are more sophisticated than the molecular
models. However, the analogy between particles and space-time events remains
a useful one.

There are many possible event-symmetric models which can be constructed
\cite{Gib95a} but the most interesting ones must be the event-symmetric
string theories \cite{Gib95b}

It might be asked what status this approach affords to events themselves.
Events are presented as fundamental entities almost like
particles. Event orientated models are sometimes known as Whiteheadian
\cite{Whi29} but Wheeler preferred to refer to a space-time viewed as a set
of events without a geometric structure as a ``bucket of dust'' 
\cite{Whe64,MiThWh73}.
In some of the models we will examine it appears as if events
are quite real, perhaps even detectable. In other models they are more
metaphysical and it is the symmetric group that is more fundamental. 
Indeed the group may only arise as a subgroup of a matrix group and the
status of an event is then comparable to that of the component of a vector.
Then again in the discrete string models we will see that events have
the same status as strings.

Above all the event-symmetric approach seems to suggest a Machian view
of physics. Space-time takes a secondary role to events which are
identified with particles or sting states. 


\section*{Discussion}

Physicists and philosophers have been interested to the small scale
structure of space-time for many years. There have been many papers
written describing various models of pregeometries or quantised
space-time and recently serious interest in such research has seen
an explosion of activity.

There have always been suggestions that space-time must be discrete
at small scales but the motivation for this assertion has changed with
time. Initially the justification was largely meta-physical. The fact
that measurement in physics does not reflect mathematical properties
of real numbers or the existence of scale dependence of physical law
could be cited as evidence.

The emergence of quantum theory led to speculations about space-time
quantisation which were reinforced by the need to renormalise in
quantum field theory. At the time many physicists found the procedure
unsatisfactory and felt that field theory could only be consistent
if there was a small scale cut-off. The measurement problem also
added to the motivation to find something more fundamental which
would manifest at small scales.

Many of these concerns have subsided but the theoretical evidence for
a minimum length at the Planck scale in quantum gravity and constraints
imposed by the black hole information loss paradox have taken their
place as motivation for discrete theories of space-time.

To systematically characterise pregeometric models we have discussed
a number of physical properties which might be either abandoned or
taken as fundamental in a pregeometry. A variety of different models
appear as a result. Some of the main classes can be summerised as
follows,

{Quantised space-time}

{Cellular Automata}

{Lattice Field Theories}

{Quantum metric spaces}

{Causal nets}

{Poset models}

{Simplicial quantum gravity}

{Topological Quantum Field Theory}

{Field theory on a cell complex}

{Non-Commutative geometry}

{Event-symmetric space-time}

In addition to these I have briefly examined the main approaches to
quantum gravity which tell us a great deal about the nature of
space-time at the smallest possible scales.

There are intricate relationships between these models which lend
hope that a realistic model of space-time may be realisable despite
the unlikelihood of direct help from experiment. Further clues on
the right way to go continue to come from string theory and 
semi-classical studies of black holes. At the same time the emergence 
of new mathematical frameworks which generalise the classical notion
of symmetry and uncover powerful relationships between algebra,
topology and field theory is at last providing us with the tools to
explore the small scale structure of space-time.

The construction of quantum groups has been absorbed into almost all
approaches to Quantum Gravity as if it was a discovery well overdue.
Discreteness of space-time is also universal but so is the importance
of topology, a clear sign that a full theory of Quantum Gravity must
resolve its discrete/continuum dual nature.

Most impressive is that old ideas such as Snyder's quantised space-time,
Regge calculus, Penrose spin networks and Wheeler's pregeometry are
now all proving to be prophetically relevant. It is a revelation of
the power of human thought that this should be the case as well as
a dramatic demonstration of the effectiveness of mathematics in physics.
An observation which must have some profound explanation.

\pagebreak


\end{document}